\documentstyle[aps,prl,preprint]{revtex}
\tightenlines
\begin{document}
\draft
\title
{ Low-Lying Energy Properties of a Frustrated Antiferromagnetic 
Spin-$\frac 1 2$ Ladder}
\author{Xiaoqun Wang }
\address{Institut Romand de Recherche Num\'erique en Physique des Mat\'eriaux,
 PPH-EPFL, CH-1015 Lausanne, Switzerland}
\date{\today}
\maketitle
\begin{abstract}
Using the Density Matrix Renormalization Group method, we determined the 
phase diagram of a frustrated antiferromagnetic spin-$\frac 1 2$ ladder at zero 
temperature. Two spin-gapped phases, the Haldane phase and the singlet phase, 
are identified. A phase transition between the two phases occurs at 
any non-zero value of frustration coupling $J_{\times}$. 
On the phase boundary, the spin gap vanishes
for sufficiently small $J_{\times}$. Crossing this non-gapped line,
the transition is of second order, while of first order for larger $J_{\times}$.
Striking frustration effects are predicted for ladder materials.

\end{abstract}
%\begin{multicols}{2}

In recent years, a lot of theoretical and experimental effort 
has been devoted to the understanding of spin-$\frac 1 2$
antiferromagnetic ladders.\cite{Rice,dago,millis,Noack,zhang,Troyer,Greven,Barnes,Az} 
These systems, which interpolate structures between one and two dimensions, 
are realized in materials such as Sr$_{n-1}$Cu$_{n+1}$O$_{2n}$\cite{Rice}.
The ladders, being arrays of coupled chains, can display significantly 
different physical properties according to their number of legs. Especially, the spin
 correlation function in the ground state shows an algebraic decay if the number
 of legs is odd and an exponential decay if it is even.\cite{Noack,Greven}
 This property essentially depends on whether or not the system exhibits a 
{\it spin gap} in the spectrum. Therefore, it seems that the ladders are 
analogous to spin-$S$ antiferromagnetic Heisenberg chains where an odd 
half-integer spin corresponds to an odd number of legs and an integer spin to an even 
number of legs.\cite{Haldane,Huse,Golinelli,Karen,Sch,Qin} 

The aim of this paper is to investigate the following two issues: i) 
How far is the above analogy valid?  In particular, are the low-lying 
energy properties of the two-leg standard ladder identical to those of the 
single $S=1$ chain? It is clear that a two-leg ladder with a ferromagnetic 
interchain coupling $J_{\bot}<0$ is equivalent to the $S=1$ chain
 when $J_{\bot} \rightarrow -\infty$. In the case of the antiferromagnetic
interchain coupling, there is also a gap opening at any $J_{\bot}$, however 
no phase transition occurs in the whole parameter range.\cite{White}. For large  $J_{\bot}$, 
the gap stems from the formation of a spin singlet on each rung. Therefore 
the origin of the spin gap for $J_{\bot}>0$ seems 
distinguishable from that in the $S=1$ chain.\cite{Rice} ii) What are the frustration
 effects caused 
by a next-nearest-neighbour interaction $J_{\times}$? Under some circumstances, 
frustration can produce either a dimerized state or a non-dimerized spin 
liquid state.\cite{afl} In the ladder material 
SrCu$_2$O$_3$,\cite{Az} the frustration is presumed to be negligible and $J_{\bot}$
comparable to the intrachain coupling $J_{\|}$.\cite{Rice} 
In this case, compared to the theoretical 
prediction of $\Delta=0.5J_{\|}$\cite{dago,Noack}, the gap is estimated 35\% smaller
by fitting the spin susceptibility and 4\% larger from the 
nuclear magnetic resonance measurement.\cite{Az} 
The origin of this discrepancy between the two estimates is unclear.\cite{Rice}
Our calculation will show that for $J_{\bot}=J_{\|}$ the spin gap only slightly changes as
$0\leq J_{\times}\leq 0.4J_{\|}$.
The maximum value of $\Delta$ is $0.519J_{\|}$ at $J_{\times}=0.25J_{\|}$. 
Therefore, it is an open question whether ladder materials are sufficiently well described 
by the standard spin ladder.

In this work, we study a two-leg frustrated antiferromagnetic 
spin-$\frac 1 2$ ladder described by:
\begin{eqnarray}
\hat H&=&\sum_{i=1,N} ~[J_{\|}(\hat {\bf S}_{1,i}\cdot\hat {\bf S}_{1,i+1}
                          +\hat {\bf S}_{2,i}\cdot\hat {\bf S}_{2,i+1})
+J_{\bot}\hat {\bf S}_{1,i}\cdot\hat {\bf S}_{2,i}
\nonumber\\
&+&~~~~~J_{\times} (\hat {\bf S}_{1,i}\cdot\hat {\bf S}_{2,i+1}
+\hat {\bf S}_{1,i+1}\cdot\hat {\bf S}_{2,i})],
\end{eqnarray}
where $\hat {\bf S}_{n,i}$ denotes a spin-$\frac 1 2$ operator at site $i$ of the 
$n$th chain. $J_{\|}$ is an intrachain coupling between two neighboring
spins in each chain, $J_{\bot}$ an interchain coupling between two spins
on each rung and $J_{\times}$ an interchain coupling between two spins of
neighboring rungs. Since $J_{\times}=J_{\|}$ is a symmetric line in the 
parameter space, we only consider the case of $J_{\times}\leq J_{\|}$. 
When $J_{\times}=0$, $\hat H$ reduces to that for 
the standard spin ladder. Hereafter we set $J_{\|}=1$.

To overcome the inherent difficulties related to the frustration, 
which plague the Monte Carlo simulation, we employ the density matrix 
renormalization group method invented by White.\cite{White2}
This method has proven to be very powerful and efficient for a systematic study 
of low-lying energy properties of low-dimensional lattice models.
In our calculation, we keep at least 240 states, and  up to 450 states are 
necessary for critical regimes. The truncation error is typically of the 
order of $10^{-8}$. Lengths up to 550 rungs are considered for open boundary 
conditions and finite size scaling is used to determine the thermodynamic limit.
For convenience, the number of rungs $N$ is chosen to be even. To check the 
accuracy and the convergence, we performed calculations keeping up to 
600 states. The errors on physical quantities are estimated to be less 
than one percent in most cases. 
The technical details will be given elsewhere.\cite{wang3} 

In Fig. 1, we show the phase diagram characterized by two fixed points: 
(i) the ``Haldane phase" so named as it contains 
the limiting case $J_{\times}=1$ and $J_{\bot}=0$, 
\begin{equation}
\hat H= \sum(\hat {\bf S}_{1,i} +\hat {\bf S}_{2,i})\cdot 
(\hat {\bf S}_{1,i+1}+\hat {\bf S}_{2,i+1})
\end{equation} 
whose low-lying spectrum is identical to that of a $S=1$ chain; 
(ii) the ``singlet phase" as it contains the case $J_{\bot}\gg 1$, 
with a ground state consisting of a singlet on each
rung and low-lying excitations generated by creating triplets on rungs.
In both cases, the system is gapped. However, a phase transition 
(shown in Figs. 3-5) occurs in the parameter space as we cross 
from (i) to (ii). 

Of special interest in the characterization of the two phases, is the   
sensitivity of their low-lying energy spectra to boundary conditions or 
impurity effects. In particular, both experiment\cite{appeli} and numerical 
calculations\cite{kennedy,wang} indicate that boundary or impurity effects
are a characteristic feature of the $S=1$ chain \cite{wang4}. 

In the {\it Haldane phase}, for open boundary conditions, 
the ground state is four-fold degenerate with total spin $S_t=0, 1$ and the continuum 
starts with $S_t=2$ as shown schematically in Fig. 2a$_{\rm o}$,
 while for periodic boundary conditions, 
the ground state becomes a singlet and the lowest excitation 
has $S_t=1$ (Fig. 2a$_{\rm p}$). We can understand this difference by considering 
the open boundary conditions as a special case of a bond 
impurity in the weak or strong coupling limit\cite{wang,wang1}. 

Furthermore, for open boundary conditions only, the region $C_1$ appears 
inside the Haldane phase, characterized by the presence of midgap states.
The number of levels is one as we enter the $C_1$ region from below (Fig. 2b) and 
grows larger and larger as we approach the phase boundary (Fig. 2c). 
This effect is a peculiar feature of this frustrated spin ladder in contrast to 
the $S=1$ chain.

In the {\it singlet phase}, independently of the boundary conditions, 
the ground state is a singlet and the continuum starts from a state 
with $S_t=1$ outside the 
region $C_2$ and $S_t=0$ inside the region $C_2$ (Fig. 2d-f). 
In contrast to the Haldane phase, neither midgap states for open 
boundary conditions nor any impurity state due to a bond impurity are found 
in this phase.\cite{wang1} 

On the {\it phase boundary}, there are at least three 
non-trivial degenerate states. Two of them have  $S_t=0$ and one has $S_t=1$.
As $J_{\times} \leq{\lambda}=0.287$, a non-gapped line shows up on the
boundary curve (Fig. 2g). Below we will discuss the phase diagram as it is 
numerically determined by the calculation of the ground state energy, 
singlet density per rung and low-lying excitations.

Fig. 3 shows the ground state energy per rung $e_0$.
For each $J_{\times}$, a phase transition between the 
Haldane phase (left) and the singlet phase (right) takes place at the 
maximum value of $e_0$.  When $J_{\times}$ is sufficiently 
large, $e_0$ is obviously singular at the critical value of
$J_{\bot}$. When $J_{\times}=1$, the singularity originates from the 
crossing of the two lowest energy levels with $S_t=0$. However, such a crossing 
{\it disappears} due to {\it level repulsion} for $J_{\times}<1$. 
The two lowest energy levels with $S_t=0$ become degenerate at the critical 
values {\it only} in the thermodynamic limit. When $J_{\times}$ is
small, the curvature of $e_0$ looks smooth at its maximum value. 
The critical values $J_{\bot,0}$ of $J_{\bot}$ are given in Table I, which 
are splined into a solid curve to indicate the phase boundary shown in Fig. 1. 
In the construction of the phase diagram, we also verified the 
transition by varying $J_{\times}$ for given $J_{\bot}$. The results 
for $J_{\bot}=1$ is shown in inset.

In order to further investigate the ground state properties, we 
calculate the singlet density per rung $\rho_s$. For a state $|\Psi\rangle$, 
we define $\rho_s=\frac 1 {N} \sum_i\langle \Psi| 
{\cal S}_i {\cal S}_i^{\dagger}| \Psi\rangle$, where ${\cal S}_i=
\frac 1 {\sqrt2}(|\uparrow_1\downarrow_2\rangle_i$$-
|\downarrow_1 \uparrow_2\rangle_i)$ is the singlet state formed 
by two spins on the $i$th rung. For $J_{\times}=1$ and in low-lying energy 
states, $\rho_s=1$ when $J_{\bot}>J_{\bot,0}$; otherwise $\rho_s=0$.
For $J_{\times}<1$, we show $\rho_s$ for the ground state in Fig. 4. 
At $J_{\bot,0}$, $\rho_s$ changes abruptly for large  
$J_{\times}$ and smoothly for small $J_{\times}$.
We note that $\rho_s=\frac 1 4$ at $J_{\bot}=J_{\times}=0$, $\rho_s<\frac 1 4$ in the 
Haldane phase, and $\rho_s>\frac 14$ in the singlet phase. Therefore,
 $\rho_s$ also characterizes all the phases.

We now discuss the low-lying excitations which are used to determine 
the regions $C_1$ and $C_2$ as well as to provide evidence for the existence 
of a non-gapped line. The spin gap $\Delta$ governs the asymptotic behavior of the 
correlation function in the ground state and the low-temperature behavior of 
the thermodynamic quantities. $\Delta$ is defined as an energy  difference between 
the ground state and the bottom state of continuum for spin systems.
In Fig. 5, we show the energies of several relevant states relative
 to the ground state energy.

For $J_{\times}=1$, the total spin on each rung is conserved. Consequently,
the states with $n$-singlets on rungs are the same as those of a 
$S=1$ chain doped by $n$ non-magnetic impurities at 
sites corresponding to those rungs at which the $n$-singlets are localized. 
In the Haldane phase, the midgap states are the lowest impurity 
states for the given number of singlets or impurities. For instance, two 
lowest midgap states shown in inset are the ground states of open 
$S=1$ chains with $N-1$ and $N-2$ spins: 
the lowest one involves a singlet at one end of an open ladder and the 
other two singlets at both ends. In the singlet phase, the low-lying 
excitations are discrete but infinitely degenerate. Two lowest 
excitations shown in inset are generated from the ground state by
 introducing one, two, and four neighbouring triplets on rungs. 
Their $S_t$ equal to one when one triplet is involved, 
or else $S_t=0$.

For $J_{\times}<1$, since the total spin on each rung is no longer conserved,
the low-lying energy properties are dramatically changed according to 
the magnitude of $J_{\times}$ and $J_{\bot}$. As shown in Fig. 5, there is no midgap 
state at $J_{\times}=0.2$ and $0.4$. For a larger $J_{\times}$, midgap states 
show up when $J_{\bot}$ is  larger than a critical value $J_{\bot,1}$.
Splining $J_{\times}$ and $J_{\bot,1}$ given in  Table I, we obtained 
a critical values for $J_{\times}$ of $0.430$.  
In the singlet phase, the infinite degeneracy at $J_{\times}=1$ is 
removed by infinitesimal $J_{\|}-J_{\times}$ so that the discrete levels 
become different bands. The bottom state has $S_t$ equal to $1$ or $0$ depnding 
on $J_{\times}$ and $J_{\bot}$.  For sufficiently large $J_{\times}$, 
the two lowest excitations with $S_t=0,1$ across at $J_{\bot}=J_{\bot,2}$. 
Several $J_{\bot,2}$ given in Table I are used to 
determine the value of $J_{\times}$, namely $\lambda=0.287$, at which the splined
curve meets the phase boundary. We note that this leads to a critical point in 
the phase diagram because the splined curve has a curvature clearly opposite 
to that of the phase boundary.

As seen from Fig. 5, for each $J_{\times}<1$ the spin gap has a minimum
given in Table I at $J_{\bot,0}^{~-}$. For sufficiently large
$J_{\times}$, $\Delta$ has a jump at $J_{\bot,0}$, e.g. see Fig. 5 
for $J_{\times}=0.8$. We found that the {\it jump} and the {\it minimum value}, 
and the {\it width} of $C_2$ decrease coherently as $J_{\times}$ becomes small
and {\it approach to zero} for $J_{\times}\rightarrow \lambda$. Therefore, 
a non-gapped line exists on the phase boundary for $0\leq J_{\times} \leq \lambda$.
On this line, $\rho_s=1/4$ and the ground state 
has a different symmetry than in the other phases.\cite{wang3} 
Further taking into account the critical behavior of $e_0$ and $\rho_s$,
we conclude that the transition between the Haldane phase and the singlet 
phase crossing the non-gapped line is of second order, while of first order for 
$J_{\times}>\lambda$. 

In Fig. 6, we show the behavior of the spin gap varying $J_{\times}$ for $J_{\bot}
=J_{\|}=1$, as is relevant to ladder materials such as SrCu$_2$O$_3$,\cite{Rice} 
It is remarkable that the spin gap is almost constant when $0\leq J_{\times}\leq 0.4$. 
At $J_{\times}=0.25$, it has a maximum value of $\Delta=0.519$.
Accordingly, it would be interesting to examine the relevance of frustration 
in ladder materials. 

We have established the phase diagram of a frustrated spin-$\frac 1 2$ ladder 
using the low-lying energy properties. Clearly, the low-lying energy properties of the 
two-leg antiferromagnetic spin-$\frac 1 2$ ladder is different from 
those of the $S=1$ chain.  According to our findings, one can naturally
expect similar effects for a frustrated $t-J$ ladder. In addition,
for very low carrier doping, bound states should occur in the Haldane gap.
As far as materials are concerned, it would be 
interesting to explore two kinds of impurity effects: 
non-magnetic and magnetic doping, which produce 
midgap states for sufficiently large value of $J_{\times}$ and 
impurity states for any non-zero value of $J_{\times}$ in the Haldane gap, 
respectively. Moreover, in order to observe the phase transition, one could measure
the specific heat and spin susceptibility on pure ladder materials for 
different interchain coupling.

The author is very grateful to  Martin Long, Felix Naef and Xenophon Zotos for 
their kind help and stimulating discussions.  
He also thanks A. Honecker, I. Peschel, and Y. Zhao for discussions.
This work is supported by the Swiss National Fond Grant No. 20-49486.96.

 \narrowtext
\begin{table}
\caption{For several values of $J_{\times}$, $J_{\bot,0},J_{\bot,1},J_{\bot,2}$ are 
the values of $J_{\bot}$ for points on the phase boundary, lower boundary of $C_1$ and 
upper boundary of $C_2$, respectively; $\Delta_{min}$ is the minimum of $\Delta$. 
Lower numbers are errors at the last digit.}
\begin{tabular}{lllllllll}
$J_{\times}$&0.2&$0.35$   &$0.4$    &$0.5   $ &$    0.6$ &$0.8    $&$1$\\
\tableline
$J_{\bot,0}$& $0.378_3$ & & $0.715_2$ & & $1.000_5$& $1.225_5$&$1.4015$\\
$J_{\bot,1}$ &&& &$0.792_8$&$0.835_5$&$0.917_5$&$0.991$  \\
$J_{\bot,2}$ & &$0.650_8$&$0.742_8$&$0.921_5$&$ 1.118_3$&$1.560_5$&$2$ \\
$\Delta_{min}$ &$0.004_4$& & $0.027_5$&& $0.130_5$ & $0.318_2$&0.4105\\
\end{tabular}
\label{table1}

\begin{figure}
\caption{Phase diagram: a thick solid curve splining dots indicates
the phase boundary; Dotted line the symmetric line;
$C_1$ is surrounded by the phase boundary, a thin curve splining diamonds and the
symmetric line; $C_2$ by the phase boundary, a thin curve 
splining squares and the symmetric line. Schematic spectra of points a--g 
are shown in Fig. 2.}
\label{fig1}  

\caption{Schematic spectra for the points a--g of Fig. 1. Lines, shaded
 rectangles and numbers denote levels, the continuum and $S_t$ of 
the levels or the continuum, respectively.}
\label{fig6}

\caption{Ground state energy per rung $e_0$ {\it vs} $J_{\bot}$ at 
$J_{\times}$$=0$,$0.2$,$0.4$,$0.6$,$0.8$, and $1$. 
In inset, it {\it vs} $J_{\times}$ at $J_{\bot}=1$. }
\label{fig2}

\caption{The singlet density per rung $\rho_s$ in the ground state {\it vs} 
$J_{\bot}$ at  $J_{\times}=0,0.2,0.4,0.6$, and $0.8$.}
\label{fig3}

\caption{Low-lying energies relative to the ground state energy {\it vs} 
$J_{\bot}$ and $J_{\times}$ for the open boundary condition. 
In the {\it Haldane phase}, solid triangles linked by solid lines
indicate the bottom state of the continuum ($S_t=2$). 
Squares linked by dashed lines the lowest midgap states (MS) ($S_t=1$). 
In the {\it singlet phase}, triangles or dots linked by solid lines and diamonds 
linked by dashed lines denote the two lowest excitations with $S_t=1,0$, 
respectively. Inset for $J_{\times}=1$.}
\label{fig4}

\caption{Spin gap {\it vs} $J_{\times}$ at $J_{\bot}=J_{\|}$}.
\label{fig5}
\end{figure}
\end{table}
% \end{multicols}

\begin{references}
\bibitem{Rice} E. Dagotto and T.M. Rice, Science, {\bf 271}, 618(1996).
\bibitem{dago} E. Dagotto, et al Phys. Rev. B {\bf 45}, 5744(1992).
\bibitem{millis} S.P. Strong aand A.J. Millis, Phys. Rev. Lett. {\bf 69}, 2419(1992).
\bibitem{Noack} S.R. White, et al Phys. Rev. Lett. {\bf 73}, 886(1994).
\bibitem{zhang} T.M. Rice, et al, Europhs. Lett. {\bf 23},
445(1993); M. Sigrist, et al, Phys. Rev. B {\bf 49}, 12058(1994).
\bibitem{Troyer} M. Troyer, et al Phys. Rev. B {\bf 50}, 13515(1994).
\bibitem{Greven} M. Greven, et al, Phys. Rev. Lett. {\bf 77}, 1865(1996).
\bibitem{Barnes} T. Barnes and J. Riera, Phys. Rev. B {\bf 50}, 6817(1994).
\bibitem{Az} M. Azuma {\it et al}, Phys. Rev. Lett. {\bf 73}, 3463(1994).
\bibitem{Haldane} F.D.M. Haldane, Phys. Lett. A{\bf 93}, 464(1983).
\bibitem{Huse} S.R. White and D.A. Huse, Phys. Rev. B {\bf 48}, 3844(1993).
\bibitem{Golinelli} O. Golinelli et al., Phys. Rev. B {\bf 50}, 3037(1994).
\bibitem{Karen} K. Hallberg, et al Phys. Rev. Lett. {\bf 76}, 4955(1996).
\bibitem{Sch} U. Schollw\"ock and Th. Jolicoeur, Europhys. Lett. {\bf 30}, 493)(1995).
\bibitem{Qin} S. Qin, et al Phys. Rev. B. {\bf 56}, R14251(1997).
\bibitem{White} S.R. White, Phys. Rev. B {\bf 53}, 52(1996).
\bibitem{afl} S.R. White and I. Affleck, Phys. Rev. B {\bf 54}, 9862(1996).
\bibitem{White2} S.R. White, Phys. Rev. Lett. {\bf 69}, 2863(1992).
\bibitem{wang3}Xiaoqun  Wang, unpublished.
\bibitem{appeli}J.F. DiTusa et al, Phys. Rev. Lett. {\bf 73}, 1857(1994).
\bibitem{kennedy} T. Kennedy, J. Phys. Condens. Matter {\bf 2}, 5737(1990).
\bibitem{wang} X. Wang and S. Mallwitz, Phys. Rev. B {\bf 48}, R492(1996).
\bibitem{wang4} 
The string order parameter alternatively distinguishes different spin-gapped phases
for an anisotropic $S=1$ chain. It is finite in the Haldane phase
and vanishes in the large-$D$ phase.\cite{str} However it is not uniquely defined
for ladders.\cite{White,wan} If the definition of Refs. \onlinecite{White,wan} is
used for the anisotropic $S=1$ chain which is maped into a frustrated ladder, 
this parameter does not vanishes in the large-$D$ phase (contradiction!). 
If the other definition is used, this parameter is finite in the Haldane 
phase and vanishes in the singlet and large-$D$ phases.\cite{wang3}
\bibitem{wang1} The nature of impurity states depends upon the manner of 
magnetic doping. Here a bond impurity means that the couplings 
$J_{\times}'\neq J_{\times}$ 
and $J_{\|}'\neq J_{\|}$ are introduced, due to local doping,
for four spins in two neighboring rungs.\cite{wang} 
The detailed results will be given elsewhere\cite{wang3}. 
\bibitem{str} M. den Nijs and K. Rommelse, Phys. Rev. B {\bf 40}, 4709(1989);
T. Kennedy and H. Tasaki, ibid {\bf 45}, 304(1992);
R.Botet, et al, ibid {\bf 28}, 3914(1983);
Y. Hatsugai and M. Kohmoto, ibid {\bf 44}, 11789(1991).
\bibitem{wan} Hiroshi Watanabe, Phys. Rev. B {\bf 52},12508(1995).
\end{references}
\end{document}